\documentclass[aps,physrev,preprint,superscriptaddress]{revtex4-2}

\usepackage{graphicx}
\usepackage{dcolumn}
\usepackage{bm}

\usepackage[utf8]{inputenc}
\usepackage[T1]{fontenc}
\usepackage{mathptmx}
\usepackage{etoolbox}
\usepackage{booktabs}
\usepackage{colortbl}
\usepackage[table]{xcolor}
\usepackage{url}

\begin{document}


\title{Self-thermometry measurements of the adiabatic temperature change in first-order phase transition magnetocaloric materials}


\author{D. O. Bastos}
\affiliation{IFIMUP - Institute of Physics for Advanced Materials, Nanotechnology and Photonics, Department of Physics and Astronomy, Faculty of Sciences, University of Porto, Rua do Campo Alegre, 687, 4169-007 Porto, Portugal}

\author{A. M. R. Soares}%
\affiliation{IFIMUP - Institute of Physics for Advanced Materials, Nanotechnology and Photonics, Department of Physics and Astronomy, Faculty of Sciences, University of Porto, Rua do Campo Alegre, 687, 4169-007 Porto, Portugal}

\author{L. Andrade}%
\affiliation{IFIMUP - Institute of Physics for Advanced Materials, Nanotechnology and Photonics, Department of Physics and Astronomy, Faculty of Sciences, University of Porto, Rua do Campo Alegre, 687, 4169-007 Porto, Portugal}

\author{R. K. Dumas}%
\affiliation{Quantum Design, Inc. 10307, Pacific Center Court, San Diego, CA 92121, 1 (858) 481-4400}

\author{J. S. Amaral}%
\affiliation{Department of Physics and CICECO - Aveiro Institute of Materials, University of Aveiro, 3810-193 Aveiro, Portugal}

\author{K. Dixon-Anderson}%
\affiliation{Division of Materials Science and Engineering, Ames Laboratory of US DOE, Iowa State University, Ames, Iowa 50011-3020, United States}
\affiliation{Department of Materials Science and Engineering, Iowa State University, Ames, Iowa 50011-2300, United States}

\author{Y. Mudryk}%
\affiliation{Division of Materials Science and Engineering, Ames Laboratory of US DOE, Iowa State University, Ames, Iowa 50011-3020, United States}
\affiliation{Department of Materials Science and Engineering, Iowa State University, Ames, Iowa 50011-2300, United States}

\author{V. Franco}%
\affiliation{Multidisciplinary Unit for Energy Science (MUFENS), Department for Condensed Matter Physics, University of Seville, PO Box 1065, 41080 Seville, Spain}

\author{J. P. Araújo}
\affiliation{IFIMUP - Institute of Physics for Advanced Materials, Nanotechnology and Photonics, Department of Physics and Astronomy, Faculty of Sciences, University of Porto, Rua do Campo Alegre, 687, 4169-007 Porto, Portugal}

\author{R. Almeida}
\email{rafael.almeida42@protonmail.com}
\affiliation{IFIMUP - Institute of Physics for Advanced Materials, Nanotechnology and Photonics, Department of Physics and Astronomy, Faculty of Sciences, University of Porto, Rua do Campo Alegre, 687, 4169-007 Porto, Portugal}

\author{J. H. Belo}
\email{jbelo@fc.up.pt}
\affiliation{IFIMUP - Institute of Physics for Advanced Materials, Nanotechnology and Photonics, Department of Physics and Astronomy, Faculty of Sciences, University of Porto, Rua do Campo Alegre, 687, 4169-007 Porto, Portugal}


\date{\today}

\begin{abstract}
Accurately measuring the magnetocaloric effect is necessary to foster the development of magnetic refrigeration devices. However, current methods are inconvenient, requiring different instruments to measure each individual property or a custom-made setup. By measuring the time-varying magnetization in a commercially available VersaLab\textsuperscript{\textregistered} PPMS\textsuperscript{\textregistered} from Quantum Design, we have determined the adiabatic temperature change ($\Delta$T$_{\textrm{ad}}$) of the first-order phase transition material Gd$_5$Si$_2$Ge$_2$, for a magnetic field change of 0 to 1 T, under high vacuum ($<$ 0.1 mTorr). For each temperature and magnetic field, the equilibrium magnetization is used as the magnetization-to-temperature conversion curve, allowing us to extend the validity of a previously proposed technique to the first-order phase transition material Gd$_5$Si$_2$Ge$_2$, which exhibits significant hysteresis. Our method thus enables full characterization (magnetic entropy change, adiabatic temperature change, and heat capacity) of any magnetocaloric material, whether it has a first-order or a second-order phase transition, using a single instrument. Comparing to a directly measured $\Delta$T$_{\textrm{ad}}$, our method resulted in a peak $\Delta$T$_{\textrm{ad}}$ value of 4.47 K, within 1\% of the directly measured value for a sample of the same composition.
\end{abstract}


\maketitle

\section{Introduction}
The refrigeration sector is responsible for 7.5\% of global greenhouse emissions and around 20\% of electricity consumption \cite{Kitanovski2020,Baha2025}, and these numbers are expected to grow as refrigeration technologies become more widely adopted in developing regions, and due to the temperature rise from climate change \cite{Programme2018,Kitanovski2020,Baha2025}. Refrigeration devices became essential to modern living, providing comfortable living spaces in homes and workplaces, and providing reliable cold supply chains for food and medicine. However, conventional vapor-compression technology, used in refrigeration and air-conditioning, relies on gases that are ozone depleting, greenhouse related, with high global warming potential (GWP), or hazardous (CFCs, HCFCs and HFCs) \cite{Programme2018,Baha2025,Kitanovski2020}. As efforts to combat climate change are needed now more than ever \cite{COP30}, researchers worldwide have been trying to improve the performance of conventional refrigeration systems and develop new refrigeration technologies.\par 

Solid state caloric cooling alternatives such as magnetocaloric \cite{Gutfleisch2011,Franco2018}, electrocaloric \cite{Shao2023}, mechanocaloric (elastocaloric \cite{Tusek2015} and barocaloric \cite{Cirillo2022}), have gained prominence due to the lack of harmful gases and potentially high efficiency when based on fully reversible effects \cite{Klinar2024,Gutfleisch2011,Franco2018}, reducing greenhouse gases emission from energy production. Magnetocaloric refrigeration remains the most widely studied alternative \cite{Klinar2024}. This type of refrigeration is based on the magnetocaloric effect (MCE), a coupling between a magnetic material's temperature and magnetization, which causes an adiabatic temperature change, $\Delta$T$_{\textrm{ad}}$, or an isothermal entropy change, $\Delta$S$_{\textrm{iso}}$, depending on speed of the magnetic field application and thermal conditions. The magnetic phase transition can be first-order or second-order. First-order phase transitions (FOPT) are due to a coupled magnetic and structural transition which are induced by an external stimulus, such as temperature or an externally applied magnetic field. Upon the phase transition, the magnetization shows a sharp and discontinuous jump, typically accompanied by hysteresis \cite{Tegus2002,Oliveira2010}. Most second-order phase transitions (SOPT), on the other hand, are purely magnetic, which leads to a smoother magnetization change without hysteresis \cite{Tegus2002,Oliveira2010}. A few methods have been proposed to distinguish between FOPT and SOPT, such as the value $\eta$ from the Bean-Rodbell model \cite{Rodbell1962,Davarpanah2018} and the n-exponent method proposed by J. Y. Law \textit{et al.} \cite{YanLaw2018}. Due to the magneto-structural coupling, FOPT materials have higher $\Delta$T$_{\textrm{ad}}$ \cite{Pecharsky2003,K.A.Gschneidner2012,Pecharsky1997,Biswas2019}, which makes them better candidates for applications in real-world devices. However, the presence of hysteresis increases the complexity of the materials' behavior, making them harder to implement and study, since their magnetization and temperature changes depend on the thermal and magnetic history of the material \cite{DiazGarcia2021,MorenoRamirez2020}.\par 

To obtain $\Delta$T$_{\textrm{ad}}$, the temperature of a sample can be directly measured before and after quickly applying a magnetic field. However, for typical lab-scale samples, this requires the use of small temperature probes, to avoid significantly deviating from adiabatic conditions \cite{Cugini2020}, motivating the use of non-contact temperature sensing techniques \cite{Revuelta2025}. Indirect measurement methods based on calorimetric data (used to estimate both $\Delta$S$_{\textrm{iso}}$ and $\Delta$T$_{\textrm{ad}}$), are also time-consuming and require specific setups and calibration techniques, especially for measurements under applied magnetic field \cite{Spichkin2014,MorenoRamirez2018}. Consequently, the mostly used method is based on a Maxwell relation, through which $\Delta$S$_{\textrm{iso}}$ is determined from bulk magnetization measurements at different temperatures and applied magnetic fields, easily done with commercial magnetometers. \par 

\begin{figure}
	\centering
	\includegraphics[width=0.5\linewidth]{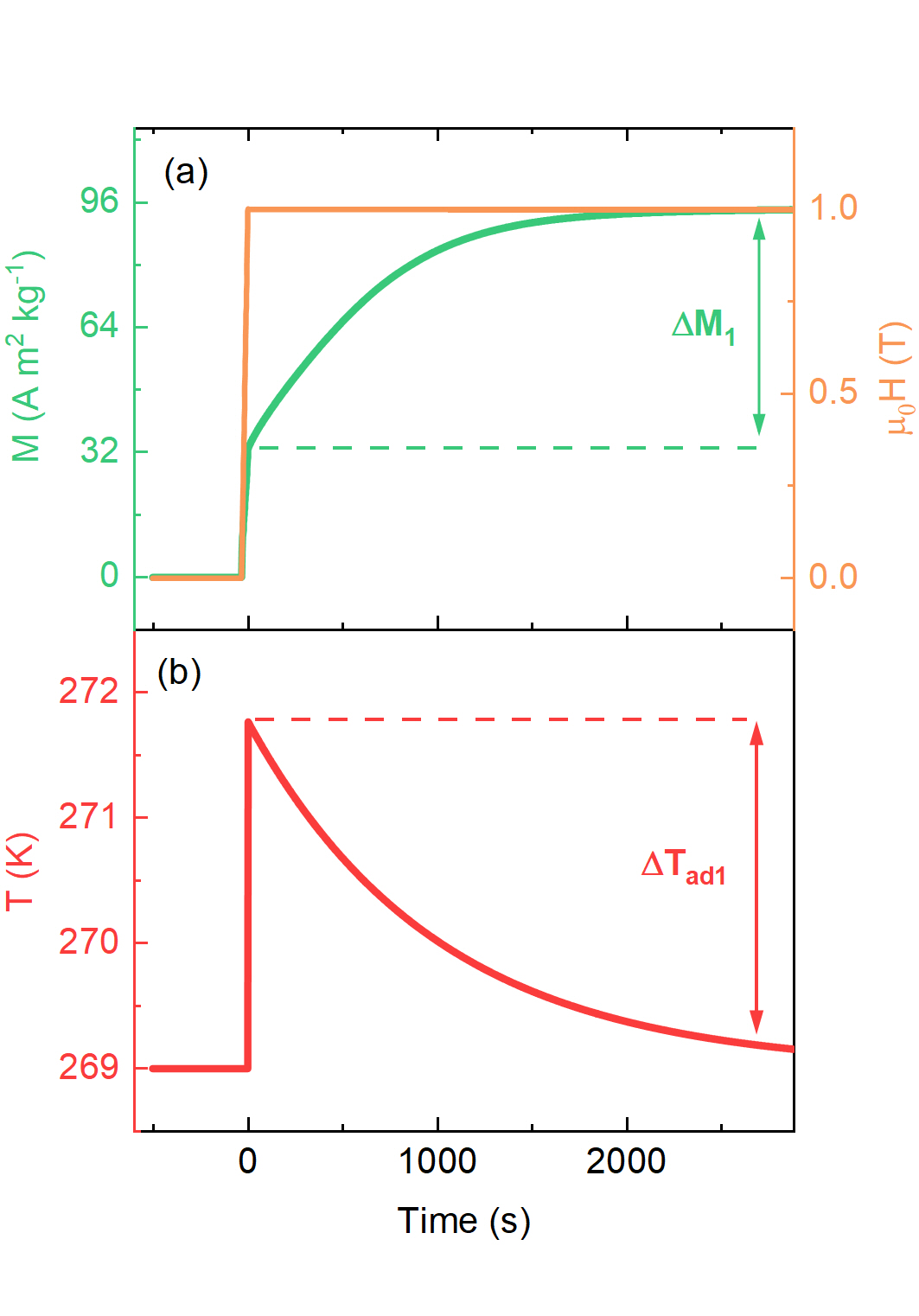}
	\caption{Schematic representation of the behavior of a magnetocaloric material during a magnetic field application. Figure (a) shows the magnetic field profile and respective magnetization variation in time, and Figure (b) shows the correspondent temperature behavior.}
	\label{MagBeh1}
\end{figure}

An alternative method has been proposed by our group to enable measuring both $\Delta$T$_{\textrm{ad}}$ and $\Delta$S$_{\textrm{iso}}$, in SOPT, using a single magnetometry device \cite{Almeida2022,Pereira2024}. These two quantities can then be used to estimate heat capacity \cite{Pereira2024}, in the region where the MCE is significant, hence covering the three most relevant physical properties to assess the magnetocaloric performance of a given material \cite{Gottschall2019,Franco2018}. The method is based on the fact that for materials exhibiting direct MCE, due to a paramagnetic to ferromagnetic phase transition from high to low temperatures, both magnetization and temperature increase when a magnetic field is applied adiabatically. After the field is stabilized, the temperature of the sample will be above the initial temperature, T$_{\textrm{i}}$, reaching T$_{\textrm{f}}=$ T$_{\textrm{i}}+\Delta$T$_{\textrm{ad}}$ and, in realistic (non-perfectly adiabatic) conditions, heat exchange will occur, leading to relaxations in temperature, with the sample cooling back to the initial temperature, and in magnetization, reaching the equilibrium value for the applied field, M$_{\textrm{eq}}$(H). This is schematized in Figure \ref{MagBeh1}. When the field is removed or decreased, the opposite behavior is obtained. By measuring both the relaxation in magnetization, M(t), and the magnetization variation with temperature, M(T), one can infer the relaxation in temperature, T(t). Although developed for materials exhibiting direct MCE, this method can easily be adapted to materials exhibiting inverse MCE. This proposed technique assumes that the magnetization \textit{versus} temperature and field response of the material is well-defined, i.e., unambiguous and non-hysteretic. This is not true for materials with FOPT. A similar technique has been proposed for a MnFe(P,Ge) compound \cite{Trung2010}, however, it neglected the above-mentioned hysteresis and focused on pulsed high magnetic fields, which are not widely available.\par 

In the current work, we expand on the method previously proposed by our group and validated in SOPT material Gd \cite{Almeida2022,Pereira2024}, to its application in FOPT materials, using Gd$_5$Si$_2$Ge$_2$ as a case study. At high temperatures, Gd$_5$Si$_2$Ge$_2$ exhibits a monoclinic structure and is paramagnetic, while at low temperature it exhibits an ortorhombic structure and ferromagnetic behavior \cite{Melikhov2015}. A decrease in temperature or an increase in magnetic field trigger a FOPT from the monoclinic paramagnetic phase to the ortorhombic ferromagnetic one.\par 

\section{Experimental Details}
\subsection{Measurement Protocol}

As mentioned above and validated in previous publications \cite{Almeida2022,Pereira2024}, for SOPT materials it is possible to obtain a magnetization-to-temperature conversion curve to allow the direct measurement of $\Delta$T$_{\textrm{ad}}$ exclusively through magnetometry. Since SOPT typically do not exhibit hysteresis, the conversion curve can be the experimentally measured magnetization \textit{versus} temperature curve, M(T), obtained at the applied field for which we intend to measure $\Delta$T$_{\textrm{ad}}$. Then, by applying the same magnetic field as quickly as possible, a $\Delta$T$_{\textrm{ad}}$ is induced in the sample (schematically represented in Figure \ref{MagBeh1}). Since the sample is no longer in thermal equilibrium with the instrument, a thermal relaxation occurs, which will be reflected in magnetization relaxation. The thermal relaxation is obtained from the measured magnetization relaxation using the magnetization-to-temperature conversion curve. The adiabatic temperature change is then obtained from the thermal relaxation.\par 

In materials that exhibit a hysteretic FOPT of the heating and cooling M(T) curves, it is not evident which curve is the most appropriate to use as the magnetization-to-temperature conversion, since the magnetization of the sample depends on its thermodynamic history. Therefore, we compared the resulting $\Delta$T$_{\textrm{ad}}$ from three conversion curve options: M(T) measured on cooling, M(T) measured on heating, and M$_{\textrm{eq}}$(T), which is constructed by pairing the average of the last 100 seconds of magnetization relaxation, M$_{\textrm{eq}}$, (to minimize the influence of noise on the data) with the sample's temperature before the magnetic field application, which should be the same as the temperature after each relaxation (see fig. \ref{MagBeh1}). We considered the temperature values read by VersaLab\textsuperscript{\textregistered}'s sample temperature sensor. It is important to note that before each measurement in high vacuum ($<$ 0.1 mTorr), the sample temperature is stabilized with a low pressure helium exchange gas present in the sample chamber to ensure proper thermalization with the nearby calibrated Cernox\textsuperscript{\textregistered} temperature sensor.\par 

We measured magnetization relaxations for a set of temperatures spanning a 250 K - 276 K range, around the sample's phase transition temperature (271 K at 1 T and 268 K at 0.05 T, determined with the M$_{\textrm{eq}}$(T) curves). Between each measurement at a given initial temperature, the sample was thermally reset by heating to 350 K, in zero field, which is far from the hysteresis temperature limits. This way, previous measurement do not affect the current measurement's results. The sample is then zero field cooled to the measurement temperature. The sample is allowed to thermalize for 15 min before high vacuum ($<$ 0.1 mTorr) is established. After another 10 min wait, the field is ramped to 1 T at 0.03 T $\textrm{s}^{-1}$, the highest sweep rate available in VersaLab\textsuperscript{\textregistered}, with the sample being kept in high vacuum to improve the adiabaticity of the process. Time = 0 s was defined as the moment at which the magnetic field reaches 1 T. The magnetic moment is then measured with the vibrating sample magnetometer (VSM) option.\par 

Since the sample is cooled from the thermal reset temperature to the measurement temperature before the magnetic field is applied, a good conversion curve candidate could be the M(T) measured on cooling. However, when the magnetic field is applied the sample heats up and then relaxes back to the initial temperature. This complex thermal history suggests that the real behavior of the material would not be fully described by M(T) on cooling but, instead, by something in between the cooling and heating M(T) curves \cite{DiazGarcia2021,MorenoRamirez2020}. As such, we may expect M$_{\textrm{eq}}$ to result in the most accurate conversions of magnetization to temperature.\par

\subsection{Samples}

A 6 g button of Gd$_5$Si$_2$Ge$_2$ was prepared through arc-melting, using high-purity gadolinium metal from the Materials Preparation Center of the Ames National Laboratory. Ge and Si were commercially sourced. The weight losses during melting were less than 1 wt.\%. The sample was annealed to improve homogeneity. The powder X-ray diffraction pattern shows pure crystallization in the monoclinic structure at room temperature \cite{Almeida2026}. A bulk sample weighing 72.10 mg, obtained from that button, was used for the indirect magnetometry measurements. It was secured with cotton inside of a gel cap, which was then adhered to a standard quartz paddle sample holder with kapton tape. The cotton also provides thermal insulation and, therefore, better adiabaticity. For the reference direct measurements, we used second piece of bulk Gd$_5$Si$_2$Ge$_2$ with m = 628.3 mg from the same button. Images of the samples and sample assemblies can be found in the supplementary material.

\subsection{Instrumentation}

All magnetometry measurements were made using a commercially available VersaLab\textsuperscript{\textregistered} PPMS\textsuperscript{\textregistered} from Quantum Design, using the VSM option and large-bore coil set. The reference direct measurements of temperature were made using a type-K thermocouple with 25 $\mu$m-diameter wire. The thermocouple was thermally coupled to the sample surface with GE-varnish and a small amount ($<$1 mg) of aluminum foil to remove excess adhesive. The sample was installed in a copper sample holder within a closed-circuit cryostat under vacuum (P$<$ 0.08 mTorr). Two thin layers of kraft paper (glued with GE-varnish) acted as a thermal resistance to slow down the heat exchange between the sample and the copper base, improving adiabatic conditions. The magnetic field was applied in under 1 s by surrounding the sample with a cylindrical Halbach array providing 1.0 T at its center. For more information regarding this setup see the supplementary material.

\section{Results and Discussion}

\begin{figure}
	\centering
	\includegraphics[width=\textwidth]{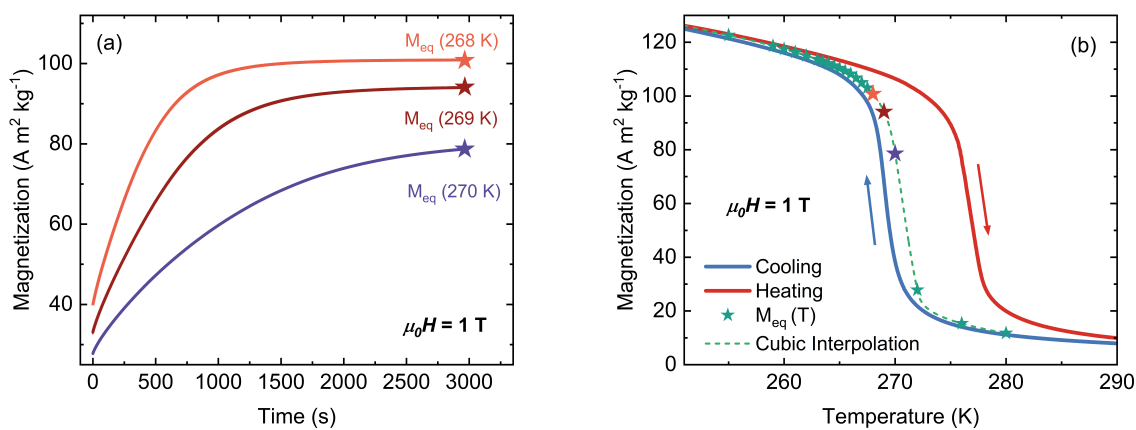}
	\caption{(a) Magnetization relaxations after a magnetic field application of 1 T, at T = 268 K, 269 K and 270 K initial temperatures, and respective equilibrium magnetizations. (b) Directly measured magnetization variation with temperature in cooling and in heating for a constant magnetic field of 1 T, and constructed M$_{\textrm{eq}}$(T) curve from magnetization relaxation data with respective cubic interpolation. The arrows near the directly measured curves indicate the direction of temperature variation during measurement.}
	\label{MT_MeqT}
\end{figure}

\begin{figure}
	\centering
	\includegraphics[width=\textwidth]{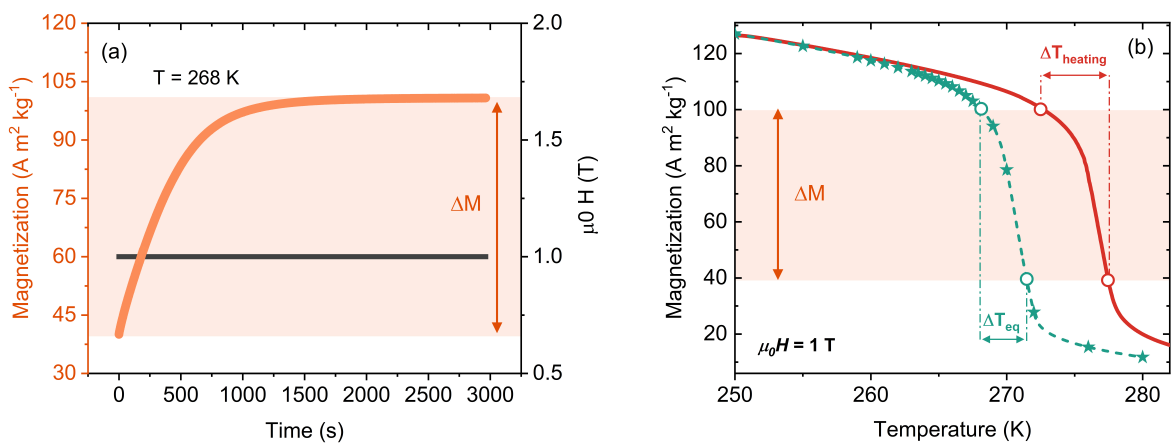}
	\caption{(a) Magnetization relaxation after a magnetic field of 1 T is applied and kept constant at a T = 268 K initial temperature. (b) Equilibrium magnetization measurements and measured magnetization variation with temperature under heating and for a constant magnetic field of 1 T. In (a) the relaxation amplitude, $\Delta$M, is schematically shown and its respective conversion to temperature shown in (b). Note how different conversion curves can lead to quite different $\Delta$T$_{\textrm{ad}}$ values.}
	\label{method_ex}
\end{figure}

Figures \ref{MT_MeqT}a and \ref{MT_MeqT}b represent, respectively, the considered M$_{\textrm{eq}}$ for a few magnetization relaxations and the measured M(T) curves compared to the constructed M$_{\textrm{eq}}$(T) curve, for a magnetic field of 1 T. As expected, due to the hysteretic behavior of the material, M$_{\textrm{eq}}$(T) lies between the heating and cooling curves. Note that it is closer to the cooling curve, which is consistent with the fact that the sample relaxes by cooling after the field application. A cubic interpolation was performed on M$_{\textrm{eq}}$(T), which gave us a smooth curve to use as the conversion. To test M(T) on cooling and on heating as a conversion curves, using a linear interpolation was enough since the curves have a high density of points. Figures \ref{method_ex}a and \ref{method_ex}b exemplify the method and show the impact of conversion curve choice on the $\Delta$T$_{\textrm{ad}}$.\par 

Figures \ref{MT_MeqT}a and \ref{method_ex}a show examples of magnetization time relaxations. The relaxation time, that is, the time that the magnetization of the sample takes to reach saturation, is quite high (in the thousands of seconds) in all cases, which is much slower than the magnetic field variation rate. This means that the sample does not have sufficient time to appreciably relax during the field application, and therefore indicates that we are, in fact, in near adiabatic conditions.\par

\begin{figure}
	\centering
	\includegraphics[width=0.6\linewidth]{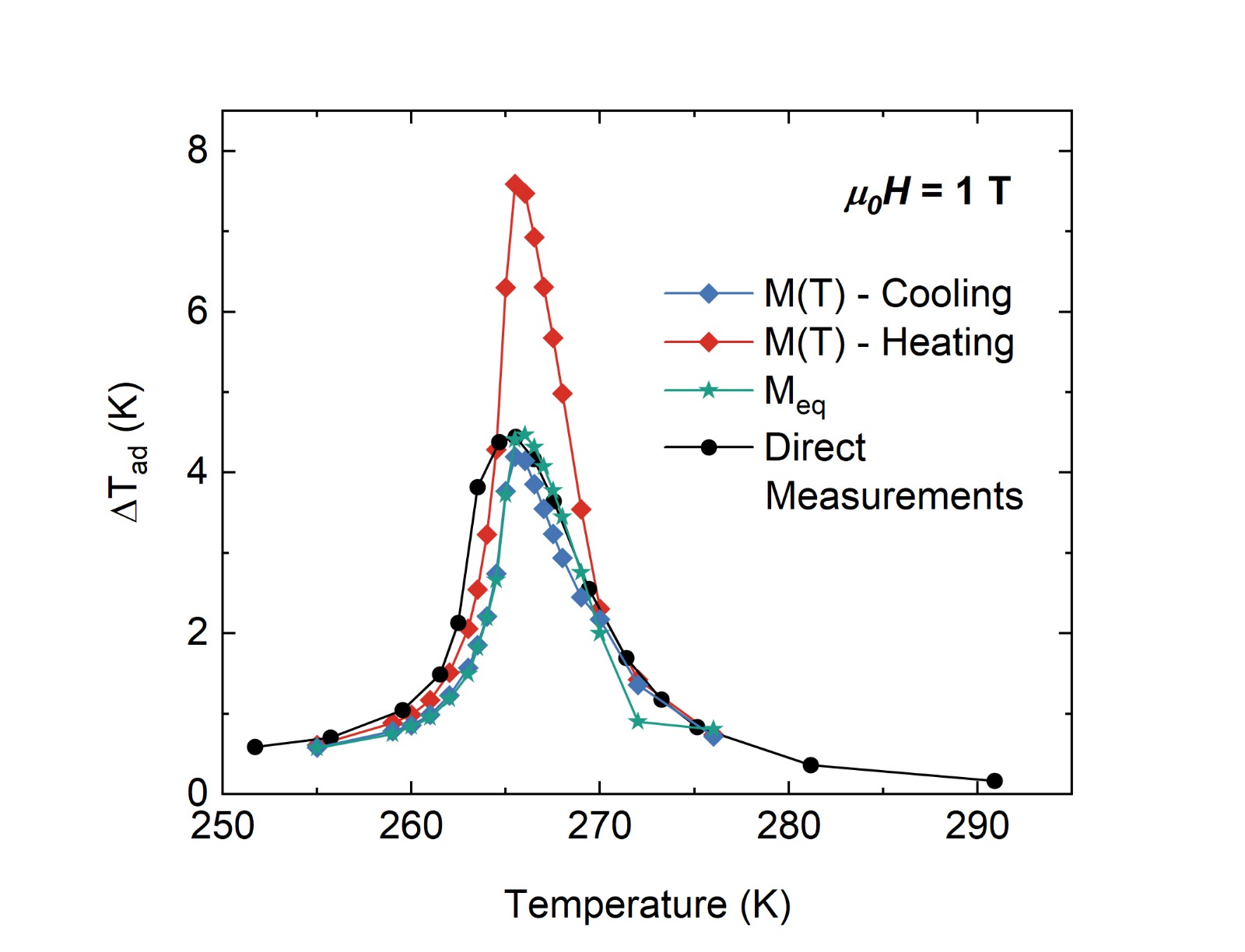}
	\caption{Results of $\Delta$T$_{\textrm{ad}}$ using the proposed method for field applications from 0 to 1 T. The different curves correspond to the different conversion curves used. A reference curve of direct measurements made with a thermocouple in a sample of the same composition is shown.}
	\label{dTad_1T}
\end{figure}

The $\Delta$T$_{\textrm{ad}}$ was determined, for each T$_\textrm{i}$, as the difference between the first and last seconds of the T(t) curve (taking the average of the last 100 seconds). Those values were used to construct a $\Delta$T$_{\textrm{ad}}$(T$_\textrm{i}$) curve, for M(T) on cooling, M(T) on heating and M$_{\textrm{eq}}$(T) as conversion curve candidates, and compare it to the $\Delta$T$_{\textrm{ad}}$ measured directly, as shown in Figure \ref{dTad_1T}. It is noticeable that M$_{\textrm{eq}}$(T) is the best conversion curve, with $\Delta$T$_{\textrm{ad}}$ values that closely align with the direct measurements. The heating M(T), on the other hand, is not an appropriate conversion curve, as it greatly overestimates the peak $\Delta$T$_{\textrm{ad}}$, as expected. During heating, the phase transition occurs at a higher temperature, causing the M(T) curve to remain smooth over a wider temperature range, and resulting in larger $\Delta$T$_{\textrm{ad}}$ peak values.\par 

Table \ref{tabTad} shows the peak value and FWHM errors and correlation coefficients of the obtained $\Delta$T$_{\textrm{ad}}$(T$_\textrm{i}$) curves in relation to the directly measured curve (see supplementary material for information on how they were determined). As it is noticeable also in Figure \ref{dTad_1T}, the $\Delta$T$_{\textrm{ad}}$ peak value is the most accurate when M$_{\textrm{eq}}$(T) is used, but the cooling M(T) produces a smaller FWHM error. However, the FWHMs of these two curves are situated where the $\Delta$T$_{\textrm{ad}}$(T$_\textrm{i}$) obtained with M$_{\textrm{eq}}$(T) gets narrower due to the point at 272 K, likely an outlier. The correlation coefficients are all similar, consistent with the fact that all $\Delta$T$_{\textrm{ad}}$(T$_\textrm{i}$) curves have similar shape, and they are close to 1. The directly measured curve is slightly broader than the others, possibly due to the larger sample size: a larger sample can have variations in chemical composition and, consequently, a T$_\textrm{c}$ distribution inside the sample, which causes a broader $\Delta$T$_{\textrm{ad}}$ peak \cite{Amaral2008,Amaral2014}. Despite this, the agreement between the directly measured curve and the indirectly obtained curve using M$_{\textrm{eq}}$ is noteworthy and corroborates the method.\par 

\definecolor{yellowgreen}{RGB}{220,255,120}
\definecolor{lightyellowgreen}{RGB}{170,255,90}

\begin{table}[]
\centering
\caption{Quantitative comparison between the $\Delta$T$_{\textrm{ad}}$(T$_{\textrm{i}}$) obtained with each conversion curve candidate and the directly measured one.}
\begin{tabular}{lccccc}
\toprule
Conversion Curve & Peak Value (K) & Peak Value Error (\%) & FWHM (K) & FWHM Error (\%) & Correlation Coefficient \\
\midrule
Directly Measured & 4.44 & -- & 7.63 & -- & -- \\
Heating M(T)      & 7.59 & \cellcolor{yellow!60}70.80 & 4.56 & \cellcolor{yellowgreen}-40.20 & \cellcolor{green!70}0.8759 \\
Cooling M(T)      & 4.20 & \cellcolor{green!60}-5.43  & 6.34 & \cellcolor{green!55}-16.90 & \cellcolor{green!70}0.8993 \\
M$_{\textrm{eq}}$(T) & 4.47 & \cellcolor{green!70}0.51 & 5.65 & \cellcolor{lightyellowgreen}-25.90 & \cellcolor{green!70}0.8707 \\
\bottomrule
\end{tabular}

\vspace{6pt}
\label{tabTad}
\end{table}

Since the FOPT results in significant hysteresis, it is also relevant to evaluate the $\Delta$T$_{\textrm{ad}}$ upon field removal and subsequent field applications (reversible $\Delta$T$_{\textrm{ad}}$), which can be drastically different \cite{Beleza2024,Cugini2020,Gottschall2019}. However, our proposed method is less effective at low magnetic fields due to the flatter conversion curve, therefore becoming more susceptible to noise. We include further details regarding this possibility in the supplementary material. Instead of measuring upon field removal, one effective way to circumvent this limitation and still obtain the reversible $\Delta$T$_{\textrm{ad}}$ would be to simply apply the magnetic field a second time (thermal reset - first field application - field removal - second field application). In this situation, our technique will presumably be as effective as demonstrated for the first field application.

\section{Conclusions}

In this work, we presented a method to indirectly determine $\Delta$T$_{\textrm{ad}}$ in FOPT Gd$_5$Si$_2$Ge$_2$, using only magnetization relaxation measurements. The $\Delta$T$_{\textrm{ad}}$(T$_\textrm{i}$) curve was obtained for this material, for a magnetic field application of 1 T, using equilibrium magnetizations, resulting from the magnetic field applications at different temperatures, as magnetization-to-temperature conversion. The obtained $\Delta$T$_{\textrm{ad}}$(T$_\textrm{i}$) curve aligns with direct measurements quite well, with a peak value error of 0.509\%, which points to the validity of the method. This work is of interest to the magnetocaloric investigation and refrigeration device engineering communities, as it demonstrates a simpler way to quantify the magnetocaloric effect in FOPT materials, using a single instrument. This was accomplished using a widely available commercial magnetometer that enables both a fast magnetic field ramp rate and the ability to measure the sample at high vacuum.\par 

Further works should aim at applying the methodology presented here to other FOPT, further extending the applicability of the procedure, and also on deepening our understanding of these materials by exploring their kinetic behavior.\par 

\section*{Supplementary Material}

See supplementary material for further details of the samples' assemblies for direct and indirect measurements, including an image that illustrates the thermal insulation used for indirect measurements. It also contains the method used to determine the correlation coefficients, peak value and FWHM, of the obtained $\Delta$T$_{\textrm{ad}}$(T$_{\textrm{i}}$) in relation to the directly measured one. Finally, the cooling and heating M(T), M$_{\textrm{eq}}$(T) and $\Delta$T$_{\textrm{ad}}$(T$_{\textrm{i}}$) curves are shown, for a magnetic field decrease of 1 T - 0.05 T, and there is brief discussion on the proposed method's lower accuracy at low fields.

\begin{acknowledgements}

The authors acknowledge FCT and its support through the projects LA/P/0095/2020, UIDB/04968/2025, and UIDP/04968/2025. R. Almeida thanks FCT for his PhD grant with reference 2022.13354.BD and DOI https://doi.org/10.54499/2022.13354.BD. This project has received funding from the European Union’s Horizon Europe research and innovation programme through the European Innovation Council under the grant agreement No. 101161135– MAGCCINE. This work was also developed within the scope of the project CICECO Aveiro Institute of Materials, UID/50011/2025 (DOI 10.54499/UID/50011/2025) \& LA/P/0006/2020 (DOI 10.54499/LA/P/0006/2020), financed by national funds through the FCT/MCTES (PIDDAC).\par 

The Gd$_5$Si$_2$Ge$_2$ sample was prepared at Ames National Laboratory. Y. Mudryk and K. Dixon-Anderson are grateful for the support from the Division of Materials Science and Engineering of the Office of Basic Energy Sciences, Office of Science of the U.S. Department of Energy (DOE). Ames National Laboratory is operated for the U.S. DOE by Iowa State University of Science and Technology under Contract No. DE-AC02-07CH11358.\par 

 V. Franco acknowledges funding from project PID2023-146047OB-I00 from AEI/10.13039/501100011033, and project PPIT2024-31833, co-financed by EU, Ministerio de Hacienda y Función Pública, FEDER and Junta de Andalucía, and VII Plan Propio de Investigación from University of Seville.

\end{acknowledgements}

\section*{Conflict of interest statement}

The authors have no conflicts to disclose.

\section*{CRediT authorship contribution statement}

\textbf{D. O. Bastos:} conceptualization, formal analysis, writing - original draft; \textbf{A. M. R. Soares:} writing - review and editing; \textbf{L. Andrade:} writing - review and editing; \textbf{R. K. Dumas:} investigation, writing - review and editing; \textbf{J. S. Amaral:} writing - review and editing; \textbf{K. Dixon-Anderson:} resources, writing - review and editing; \textbf{Y. Mudryk:} resources, writing - review and editing; \textbf{V. Franco:} writing - review and editing; \textbf{J. P. Araújo:} funding acquisition, writing - review and editing; \textbf{R. Almeida:} conceptualization, writing - review and editing; \textbf{J. H. Belo:} conceptualization, supervision, funding acquisition, writing - review and editing.

\section*{Data Availability Statement}

The data that support the findings of this study are available from the corresponding author upon reasonable request.

\nocite{*}
\bibliography{apssamp_ascii}

\end{document}